\documentstyle [prl,multicol,aps,psfig]{revtex}
\begin{document}
\title{Effect of the measurement on the decay rate of a quantum system}
\author{Brahim Elattari$^{1,2}$ and S.A. Gurvitz $^3$\\
$^1$Weizmann Institute of Science, Department of Condensed Matter
    Physics\\ 76100 Rehovot, Israel\\
$^2$Universit\'e Choua\"\i b Doukkali, Facult\'e des Sciences,
El Jadida, Morocco\\
$^3$Weizmann Institute of Science, Department of Particle 
    Physics\\ 76100 Rehovot, Israel} 
\vspace{18pt}
\maketitle

\begin{abstract}
We investigated the electron tunneling out of a quantum dot 
in the presence of a continuous monitoring by a detector. 
It is shown that 
the Schr\"odinger equation for the whole system can be reduced to new 
Bloch-type rate equations describing the time-development of the detector 
and the measured system at once. Using these equations we find that 
the continuous measurement of the unstable system does not affect 
its exponential 
decay, $\exp (-\Gamma t)$, contrary to expectations based on 
the Quantum Zeno effect . However, the width of the energy distribution 
of the tunneling electron is no more $\Gamma$, but increases due to 
the decoherence, generated by the detector. 
\end{abstract}
\hspace{1.5 cm}  PACS: 73.23.Hk.03.65.Bz.73.23.-b
\vspace{18pt}
\begin{multicols}{1}
It was suggested that an unstable quantum system slows down its decay rate
under frequent or continuous observations\cite{zeno}. This phenomenon,  
known as the quantum Zeno effect, is believed to be related to  
the projection postulate in the theory of quantum measurements\cite{neu}.
Indeed, in the standard example of two-level systems, the probability of 
a quantum transition from an initially occupied 
unstable state is $Q(\Delta t)=a(\Delta t)^2$.  
If we assume that $\Delta t$ is the measurement time, which 
consists in projecting the system onto the initial state, 
then after $N$ successive measurements the probability of 
finding the unstable system in its initial state, at time $t=N\Delta t$, is 
$P(t)=[1-a(\Delta t)^2]^{(t/\Delta t)}$. 
It follows from this result that $P(t)\to 1$ for $\Delta t\to 0$, 
i.e. suppression of quantum transition.  

Originally quantum Zeno effect has been considered as 
a slowing down of the decay rate \cite{zeno} of quantum systems in which a 
discrete initial state is coupled to a continuum of final states. This 
coupling leads to an irreversible exponential decay from 
the discrete state to the continuum of states. 
This situation is very often encountered in physics, as for instance,  
the $\alpha$--decay of a nucleus, the 
spontaneous emission of a photon by an excited atom, the photoelectric 
effect, and so on. 
But from the theoretical and experimental point of view, 
the effort has been mainly concentrated on quantum transitions 
between isolated levels\cite{zeno1} characterized by an oscillatory behavior 
between the different states. In this latter case the slowing down 
of the transition rate has, indeed, been found. However, 
this was attributed to the decoherence generated by the detector 
without an explicit involvement of the projection 
postulate\cite{stod,gur1}. On the other hand, the slowing down of the 
exponential decay rate still remains a controversial issue, despite the 
fact that it is extensively studied\cite{Fonda,schul,kof,Panov} 
and further investigations are clearly desirable.  
 
In this letter we focus our attention on the quantum Zeno effect 
in exponentially decaying systems, using a microscopic description   
which includes the measurement devices. The latter 
is an essential point missed in many studies. This work is motivated by 
the distinct difference between continuum energy levels
and discrete levels, stated above, as well as by 
the theoretical and experimental importance of the subject. 
We also propose an experimental set up which is within reach 
of nowadays experimental techniques and within which the quantum 
Zeno effect for exponentially decaying systems can be investigated. Our 
results showed that while the decay rate of the quantum unstable system is
unaffected by the measurement the energy distribution of the emitted particles 
can be strongly affected.    
\begin{figure} 
\centering{\psfig{figure=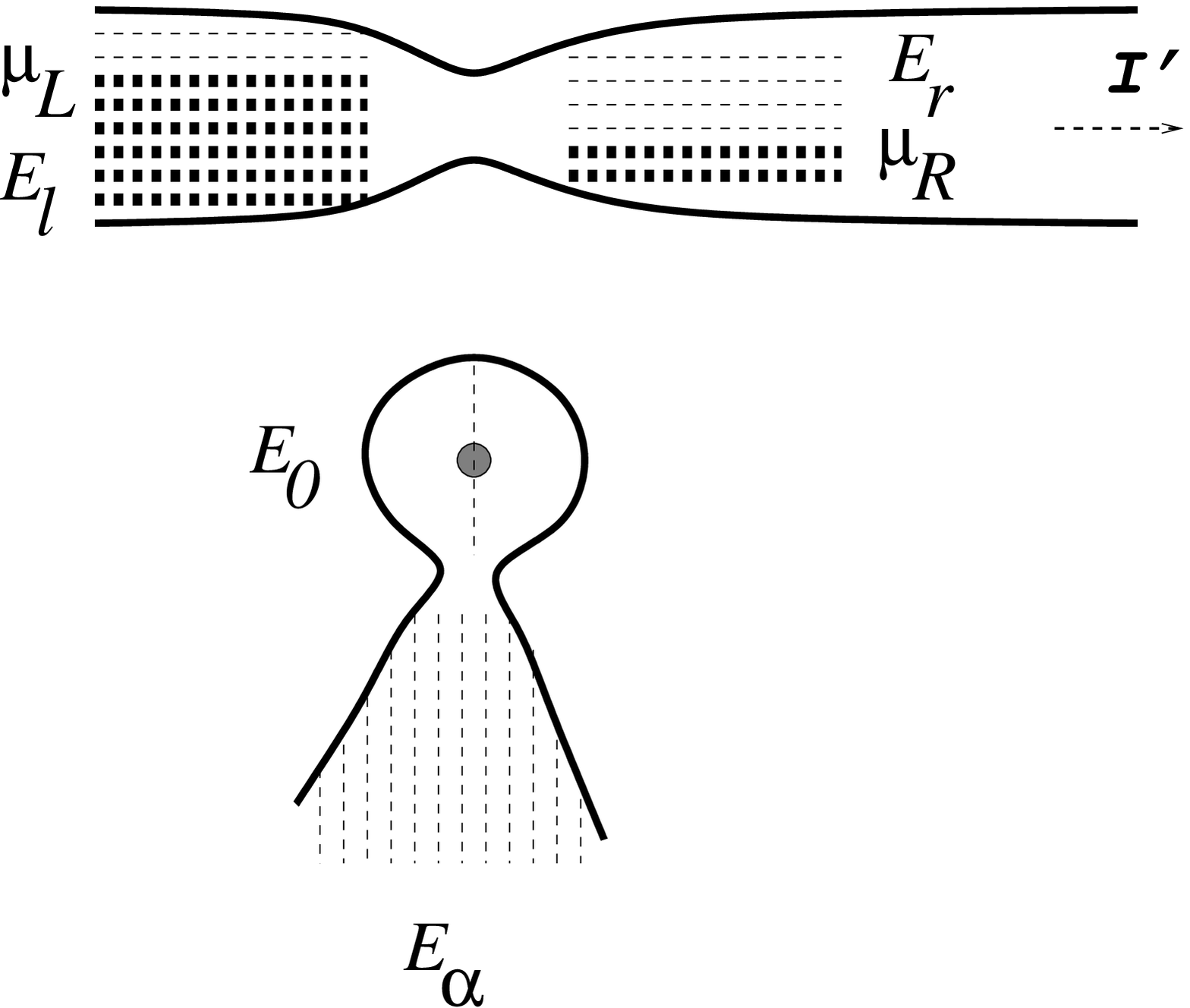,height=6cm,width=5cm,angle=0}}
{\bf Fig.~1:}
The point--contact detector near the quantum dot. 
The energy level $E_0$ of the dot is occupied by an electron,
which tunnels to continuous states $E_\alpha$ of the reservoir. $\mu_L$ and 
$\mu_R$ are the Fermi levels in the emitter and the collector, respectively. 
\end{figure}
Let us consider an electron tunneling out of a 
quantum dot to a reservoir of very dense (continuum) 
states, $E_\alpha$. The dot is placed near a quantum point-contact 
connected with two separate reservoirs (Fig. 1). 
The reservoirs are filled up to the 
Fermi levels $\mu_L$ and $\mu_R$, respectively.  Therefore the 
current $I=e^2TV/2\pi$ flows from the left (emitter) to the right 
reservoir (collector),
where $T$ is the transmission coefficient of the point-contact
and $eV=(\mu_L-\mu_R)$ is the bias voltage\cite{land}. 
(We consider the case of zero temperature). However, 
when the dot is occupied, Fig.~1, the transmission coefficient of
the point-contact decreases ($T'<T$) due to Coulomb repulsion generated by 
the electron inside the dot. Respectively, the current through   
the quantum dot diminishes, $I'<I$. Thus, the point-contact does  
represent a detector, which monitors the occupation of the quantum dot.
Actually, such a point-contact detector has been successfully used 
in different experiments \cite{Buks}.  
Notice that the current variation ($I-I'$) can be a macroscopic
quantity if the applied voltage $V$ is large enough. 
The dynamics of the entire system is determined by 
the many-body time-dependent Schr\"odinger equation 
$i|\dot\Psi(t)\rangle=H|\Psi(t)\rangle$, 
where the total Hamiltonian consists of three components 
$H=H_{QD}+H_{PC}+H_{int}$,
describing the quantum dot, the point-contact detector, and 
their mutual interaction, respectively.
These three parts can be written in the form of tunneling Hamiltonians, as 
\begin{eqnarray}
H_{QD}&=&E_0c_0^\dagger c_0+\sum_\alpha E_\alpha c_\alpha^\dagger c_\alpha
+\sum_\alpha [\Omega_\alpha c_\alpha^\dagger c_0+H.c.] ,
\nonumber\\
H_{PC}&=&\sum_l E_l c_l^\dagger c_l+\sum_r E_r c_r^\dagger c_r
+\sum_{l,r} [\Omega_{lr} c_l^\dagger c_r+H.c.]
\nonumber\\
H_{int}&=&\sum_{l,r} [\delta\Omega_{lr} c_l^\dagger c_r+H.c.]c_0^\dagger c_0
\label{a1}
\end{eqnarray}
Where the operators $c_{i}^+ (c_{i})$ correspond to the creation 
(annihilation) of an electron in state $i$. 
The $\Omega_\alpha$ and $\Omega_{lr}$ are the hopping 
amplitudes between the
states $E_0$, $E_\alpha$ and $E_l$, $E_r$ respectively.  
These amplitudes are found to be directly 
related to the tunneling rate of the electron out of the quantum dot ($\Gamma$) 
and to the penetration coefficient of the point-contact ($T$) as 
$\Gamma =2\pi |\Omega_\alpha|^2\rho$ and $T=(2\pi )^2|\Omega_{lr}|^2
\rho_L\rho_R$, respectively. Here $\rho$ are the density of states in the 
corresponding reservoirs.  The quantity 
$\delta\Omega_{lr}=\Omega'_{lr}-\Omega_{lr}$ represents the variation 
of the point-contact hopping amplitude, when the dot is occupied. 
In our derivations we assume that $\Omega$ and $\rho$  are weakly 
energy dependent and $(\mu_L-\mu_R)\gg \Omega^2\rho$. The latter condition 
is necessary for the exact solubility of the model.
    
Consider the entire system in the initial condition, corresponding 
to occupied quantum dot and filled reservoirs up to Fermi levels $\mu_L$ 
and $\mu_R$, Fig.~1, denoted by 
$|\Psi (0)\rangle =|0\rangle$. This state 
is not stable: the Hamiltonian (\ref{a1}) requires it to decay to continuum
states having the form $c^\dagger_ic^\dagger_{i_1}\cdots c_jc_{j_1}
\cdots |0\rangle$.
In general, the total
wave-function at time $t$ can be written as 
\begin{eqnarray}
&&|\Psi (t)\rangle = \left [ b_0(t) 
+\sum_{l,r} b_{lr}(t)c_r^{\dagger}c_l+\sum_\alpha b_\alpha(t)
c_\alpha^\dagger c_0
\right.\nonumber\\
&&\left. ~~~~~~~~~~~~~~~~~~~~~~~~ 
+\sum_{\alpha ,l,r} b_{\alpha lr}(t)c_r^{\dagger}c_\alpha^{\dagger}c_lc_0
+\cdots\right ]|0\rangle\ .
\label{a2}
\end{eqnarray}
Where $b(t)$ are the probability amplitudes of finding the system 
in the state
defined by the corresponding creation and annihilation operators.
Using these amplitudes one finds the reduced 
density-matrix by tracing out the irrelevant degrees of freedom. 
This density matrix will give us all probability distributions
describing the behavior of the entire system. For instance,
$\sigma^{(0)}_{00}(t)=|b_0(t)|^2$, is the 
probability of finding the system in the initial state at time $t$, 
$\sigma^{(1)}_{00}(t)= \sum_{l,r}|b_{lr}(t)|^2$ is the probability of
finding one electron in the collector and the quantum dot is occupied,
$\sigma^{(0)}_{\alpha\alpha}(t)=\sum_\alpha |b_\alpha (t)|^2$ is the 
 probability for the electron to tunnel out of the dot into 
 level $E_\alpha$ and 
no electron arriving the collector, and so on. In general,
the total probability for the electron to occupy the dot is
$\sigma_{00}(t)=\sum_n\sigma^{(n)}_{00}(t)$, and the probability 
of tunneling into level $E_\alpha$ is $\sigma_{\alpha\alpha}(t)=
\sum_n\sigma^{(n)}_{\alpha\alpha}(t)$. Here the subscript $n$ denotes 
the number of electrons reaching the collector at time $t$.   
The corresponding off-diagonal density-matrix elements 
$\sigma_{0\alpha}=\sigma^*_{\alpha 0}$ describe the electron 
in the linear superposition of the states $E_0$ and $E_\alpha$.

In order to find the amplitudes $b(t)$, we 
substitute Eq.~(\ref{a2}) into the time dependent 
Scr\"odinger equation and use the
Laplace transform $\tilde b(E)=\int_0^\infty b(t)\exp (iEt)dt$.
Then we find an infinite set of algebraic equations for the amplitudes 
$\tilde b(E)$, given by 
\begin{mathletters}
\label{a3}
\begin{eqnarray} 
&&E\tilde b_0-\sum_\alpha\Omega_\alpha\tilde b_\alpha 
-\sum_{l,r}\Omega'_{lr}\tilde b_{lr}=i\ ,
\label{a3a}\\
&&(E+E_0-E_\alpha)\tilde b_\alpha -\Omega_\alpha\tilde b_0
-\sum_{l,r}\Omega_{lr}\tilde b_{\alpha lr}=0\ ,
\label{a3b}\\
&&(E+E_l-E_r)\tilde b_{lr}-\sum_\alpha\Omega_\alpha\tilde b_{\alpha lr}
\nonumber\\
&&~~~~~~~~~~~~~~~~~~~
-\Omega'_{lr}\tilde b_0-\sum_{l',r'}\Omega'_{l'r'}\tilde b_{ll'rr'}=0
\label{a3c}\\
&&~~~~~~~~~~~~~~~~~~~~~~\cdots\cdots\cdots 
\nonumber
\end{eqnarray}
\end{mathletters}
It is very important that the tracing of the reservoir variables 
can be carried out directly in Eqs.~(\ref{a3}) without their explicit 
solutions. As a result, Eqs.~(\ref{a3}) are converted to
the Bloch-type equations for the reduced density-matrix 
$\sigma_{ij}^{(n)}(t)$. Such a technique has been derived in
\cite{gur1,gur2}. In this paper we generalize it 
by converting Eqs.~(\ref{a3}) into rate equations without tracing  
over all the continuum states. We, thus, obtain generalized 
Bloch-type equations which determine the energy distribution of the 
tunneling particles. In the following we outline this derivation,
relegating the technical details to a more extended publication. 

First, we replace each of the sums in Eqs.~(\ref{a3}) by an   
integral, $\sum_k \to \int \rho (E_k)dE_k$   
 which can be treated analytically. The Eq.~(\ref{a3a}), for instance, 
after replacing in it the amplitudes $\tilde b_\alpha$ and 
$\tilde b_{lr}$ by the corresponding expressions obtained   
from Eqs.~(\ref{a3b}), (\ref{a3c}), becomes 
\begin{eqnarray}
&&\left (E-\int {\Omega_\alpha^2\rho (E_\alpha)dE_\alpha
\over E+E_0-E_\alpha}\right.\nonumber\\
&&\left.-\int{{\Omega'}^2_{lr}\rho_L(E_l)\rho_R(E_r)dE_ldE_r
\over E+E_l-E_1-E_r}
\right )\tilde b_0(E)+{\cal F}=i\ ,  
\label{a33}
\end{eqnarray}
where ${\cal F}$ denotes the terms in which the amplitudes $\tilde b$ 
cannot be factorized out the integrals. These terms vanish when the 
integration limits are extended to infinity (the large bias limit, 
$(\mu_L-\mu_R)\gg \Omega_{L,R}^2\rho_{L,R}$). This is due to the fact that the 
singularities of the amplitudes $\tilde b$, as functions of the variables  
$E_{l,r,\alpha}$, lie on the same side of the integration contour. 
The remaining integrals in Eq.~(\ref{a33}) can be splited into a sum of  
singular and principal parts. The singular parts yield  $i\Gamma /2
+iD'/2$, where $D'=(2\pi )|\Omega'_{lr}|^2\rho_L\rho_R(\mu_l-\mu_R)$.
While the principal parts induce a shift of energy which is merely 
absorbed by a redefinition of the energy 
levels. Performing the same procedure with all other equations (\ref{a3}),   
we reduce them to the following system of equations for the amplitudes 
$\tilde b(E)$\cite{gur1,gur2} 
\begin{mathletters}
\label{a4}
\begin{eqnarray} 
&&\left (E+i{D'\over 2}+i{\Gamma \over 2}\right )\tilde b_0(E)=i\ ,
\label{a4a}\\
&&\left (E+E_0-E_\alpha +i{D\over 2}\right )\tilde b_\alpha (E) 
-\Omega_\alpha\tilde b_0(E)=0\ ,
\label{a4b}\\
&&\left (E+E_l-E_r+i{D'\over 2}+i{\Gamma \over2}\right )\tilde b_{lr}(E) 
-\Omega'_{lr}\tilde b_{lr}(E)=0
\label{a4c}\\
&&~~~~~~~~~~~~~~~~~~~\cdots\cdots\cdots 
\nonumber
\end{eqnarray}
\end{mathletters}
where $D=(2\pi )|\Omega_{lr}|^2\rho_L\rho_R(\mu_l-\mu_R)$.
In order to transform Eq.~(\ref{a4}) to equations for the density-matrix  
we multiply each of them by the corresponding complex conjugate 
amplitude $b^*(E')$. 
For instance by multiplying Eq.~(\ref{a4b}) by $b_\alpha^*(E')$ and 
subtracting its complex conjugated equation multiplied by $b_\alpha(E)$
we obtain 
\begin{eqnarray}  
&&(E-E'+iD)b_\alpha(E)b_\alpha^*(E')\nonumber\\
&&~~~~~~~=\Omega_\alpha
[b_0(E)b_\alpha^*(E')-b^*_0(E')b_\alpha(E')]
\label{a6}
\end{eqnarray}
It is quite easy to see that the inverse Laplace transform turns this 
equation to the following one for the density-matrix
\begin{equation} 
\dot{\sigma}_{\alpha\alpha}^{(0)} =  -D\sigma_{\alpha\alpha}^{(0)}
+i\Omega_\alpha(\sigma_{0\alpha}^{(0)}-\sigma_{\alpha 0}^{(0)})
\label{a7}
\end{equation}

Proceeding in the same way with all other equations (\ref{a4}) and 
integrating over the continuum states of the collector and the emitter, we 
obtain the following infinite set of equations for the density matrix 
$\sigma (t)$
\begin{mathletters}
\label{Reqn}
\begin{eqnarray}
\label{Reqn1}
\dot{\sigma}_{00}^{(n)} & = & -(\Gamma+D')\sigma_{00}^{(n)}
+D'\sigma_{00}^{(n-1)}\\
\label{Reqn2}
\dot{\sigma}_{\alpha\alpha}^{(n)} & = & -D\sigma_{\alpha\alpha}^{(n)}+
D\sigma_{\alpha \alpha}^{(n-1)}+i\Omega_\alpha(\sigma_{0\alpha}^{(n)}
-\sigma_{\alpha 0}^{(n)})\\
\label{Reqn3}
\dot{\sigma}_{\alpha 0}^{(n)} & = & i(E_0-E_\alpha)\sigma_{\alpha 0}^{(n)}
-i\Omega_\alpha \sigma_{00}^{(n)}-\frac{\Gamma+D+D'}{2}\sigma_{\alpha 0}^{(n)}
\nonumber\\
&&~~~~~~~~~~~~~~~~~~~~~~~~~~~~~~~~~~~~~+\sqrt{DD'}\sigma_{\alpha 0}^{(n-1)}.
\end{eqnarray} 
\end{mathletters}
Eqs.~(\ref{Reqn}) are a generalization of the 
previously derived Bloch-type rate-equations for quantum transport 
in mesoscopic systems\cite{gur1,gur2}. They have a clear physical 
interpretation. 
Consider for example Eq.~(\ref{Reqn1}) for the probability 
of finding the electron inside the dot and $n$ electrons in the collector. 
It decreases due to one-electron hopping to the collector 
(with rate $D'$), or due to the electron tunneling out of the dot 
(with rate $\Gamma$). These processes are described by the first 
(``loss'') term in Eq.~(\ref{Reqn1}). On the other hand, there exists 
the opposite (``gain'') process when the state with 
$(n-1)$ electrons in the collector converts into the state with $n$ 
electrons in the collector. It also takes place 
due to penetration of one electron through the point-contact with the same 
rate $D'$ (the second term in Eq.~(\ref{Reqn1})).  

The evolution of the off-diagonal density-matrix elements 
is given by Eq.~(\ref{Reqn3}). 
It can be interpreted in the same way as the rate equation for the 
diagonal terms. Notice, however, the difference between the ``loss'' 
and the ``gain'' terms. The latter can appear only 
due to coherent transition of the whole linear 
superposition\cite{gur1,gur2}. 

Since our rate equations distinguish between different 
continuum states ($E_\alpha$), we can find the energy 
distribution of the tunneling electron by tracing out
the detector states $n$ in Eqs. (\ref{Reqn}). 
As a result we obtain the following final equations 
for the electron density-matrix $\sigma (t)$
\begin{mathletters}
\label{eqn}
\begin{eqnarray}
\dot{\sigma}_{00} & = & -\Gamma\sigma_{00}
\label{eqn1}\\
\dot{\sigma}_{\alpha\alpha} & = & i\Omega_\alpha(\sigma_{0\alpha}
-\sigma_{\alpha 0})
\label{eqn2}\\
\dot{\sigma}_{\alpha 0} & = & i(E_0-E_\alpha )\sigma_{\alpha 0}-
i\Omega_\alpha \sigma_{00}-\frac{\Gamma+\Gamma_d}{2}\sigma_{\alpha 0}\ .
\label{eqn3}
\end{eqnarray}
\end{mathletters}
Here $\Gamma_d=(\sqrt{D}-\sqrt{D'})^2$ is 
the decoherence rate\cite{gur1}.

It is instructive to compare Eqs.~(\ref{eqn}) 
with the similar Bloch-type equations describing quantum transitions
between two isolated levels\cite{gur1,hacken}. 
In the case of isolated levels ($E_1$ and $E_2$), the equations for
the density-matrix $\sigma$ are symmetric with respect to $E_1$ and $E_2$. 
Whereas in the case of transition between the isolated 
($E_0$) and the continuum states ($E_\alpha$)  
the corresponding symmetry, between $E_0$ and $E_\alpha$, is broken    
as can be seen, for example, in the equation for the off-diagonal 
term $\sigma_{0\alpha}$ where the coupling 
with $\sigma_{\alpha\alpha}$ is missing. Eq.~(\ref{eqn3}).

The probability of finding the electron inside the dot is 
obtained directly from Eq.~(\ref{eqn1}) given by 
\begin{equation}
\sigma_{00}(t)=\exp\ (-\Gamma t).
\label{dec}
\end{equation}
It means that the continuous monitoring 
of the unstable system does not slow down its exponential decay. 
Nevertheless, it can be shown that  
the energy distribution of the tunneling electron,
$P(E_\alpha)\equiv \sigma_{\alpha\alpha}(t\to\infty )$, is affected. 
Indeed, by solving Eqs.~(\ref{eqn}) in the limit of $t\to\infty$ we find 
a Lorentzian distribution centered about $E_\alpha=E_0$:
\begin{equation}
P(E_\alpha)=\frac{|\Omega_\alpha|^2}
{\Gamma}\frac{\Gamma+\Gamma_d}{(E_0-E_\alpha)^2+(\Gamma+\Gamma_d)^2/4} .
\label{a5}
\end{equation}
If there is no coupling with the detector, $\Gamma_d=0$, 
the Lorentzian width (the line-width) $\Gamma$ is exactly  
the inverse life-time of the quasi-stationary state, Eq.~(\ref{dec}).
However, it follows from Eq.~(\ref{a5}) that the measurement results  
in a broadening of the line-width, which becomes  
$\Gamma+\Gamma_d$ due to the decoherence generated by the detector. 
At first sight this result might look very surprising. Indeed, it
is commonly accepted that the line-width does correspond 
to the life-time. Yet, we demonstrated here that it might not be the 
case when the system interacts with an environment (the detector). 
To understand this result, one might think of the following argument. 
Due to the measurement, the energy level $E_0$ suffers an additional 
broadening of the order of  $\Gamma_d$. However, this broadening does 
not affect the decay rate of the electron $\Gamma$, 
since the exact value of $E_0$ relative to $E_\alpha$ is irrelevant 
to the decay process. In contrast, the probability distribution $P(E_\alpha)$ 
is affected because it does depend on the position of $E_0$ relative to 
$E_\alpha$ as it can be seen in
Eq.~(\ref{a5}).

Although our result has been proved for a specific detector, we expect 
it to be valid for the general case, provided that the density of 
states $\rho$ and the transition amplitude $\Omega$ vary slowly with 
energy. This condition is sufficient to obtain a pure exponential decay 
of the state $E_0$.  On the contrary, if $\rho_\alpha (E_\alpha )$ or 
$\Omega_\alpha$ 
depend sharply on energy, then the integrals in Eq.~(\ref{a33}) 
yield additional $E$-dependent terms
that modify both the exponential dependence of the decay
probability, Eq.~(\ref{dec}) and the energy distribution, Eq.~(\ref{a5}). 
As a result, the measurement process could hinder the decay rate\cite{BG}.

We emphasize that our results were obtained from the Schr\"odinger equation
describing the dynamical evolution of the entire system, 
without explicit use of the projection postulate. 
This is in contrast with other works, as for instance\cite{Fonda},
where the reduction was repeatedly involved during the continuous 
measurement process. Although our final result does not display any  
slowing down of the decay rate, it should not be considered 
as a contradiction with the projection 
postulate. Indeed, the hindering of the decay rate, generated by the
projection postulate, relies on the assumption that the probability 
of transitions between different quantum states  
is $Q(\Delta t)=a(\Delta t)^2$.
It is definitely correct for transitions between isolated states,
where the transition probability has an oscillatory behavior. However, 
in the case of transitions from isolated 
to very dense states, $Q(t)$ represents 
a sum of many oscillations with close frequencies. When  
averaged over the time-interval 
$\delta t=1/\bar{\cal E}$, where $\bar{\cal E}$ is the width of the function 
$|\Omega(E)|^2 \rho(E)$  (in our case $\bar{\cal E}\to\infty$),  
the resulting $Q(t)$ would then represent a pure exponential decay. 
In this case $Q(\Delta t)\propto \Delta t$, so that repeated applications  of 
the projection postulate would not change the life-time of the decayed state.
But, if $\delta t$ is finite that zero, due to the energy dependence of $\rho$ 
or $\Omega$, then $\sigma_{00}(t)$ exhibits deviations from the exponential 
behavior, which could result in quantum Zeno effect \cite{Fonda}.
  
In conclusion, we have given a microscopic description of quantum Zeno effect
in pure exponentially decaying quantum systems including the 
measurement devices. 
Our results show that while the measurements does not affect the decay rate,
the energy distribution of the tunneling electron is broadened. This
description applies to a wide range of physical processes as mentioned at the 
beginning of this letter.  
In particular, it can be verified in experiments with mesoscopic 
quantum dots, by using the point-contact detector\cite{Buks}, 
or an alternative set up.   

We thank S. Levit and Y. Imry for most valuable discussions.
One of us (S.G.) would like to acknowledge the hospitality of 
Oak Ridge National Laboratory and 
TRIUMF, while parts of this work were being performed. One of us 
(B.E.) gratefully acknowledge the support of GIF and the
Israeli Ministry of Science and Technology and the French Ministry of
Research and Technology

\end{multicols}
\end{document}